\documentclass[letterpaper,aps,prb,twocolumn,showpacs,superscriptaddress]{revtex4}

\usepackage{graphicx}

\begin{document}

\title{Magnetic properties of
$\mathbf{Ni}_{2.18}\mathbf{Mn}_{0.82}\mathbf{Ga}$ Heusler alloys
with a coupled magnetostructural transition}

\author{V.~V.~Khovailo}
\author{T.~Takagi}
\author{J.~Tani}
\affiliation{Institute of Fluid Science, Tohoku
University, Sendai 980--8577, Japan}

\author{R.~Z.~Levitin}
\affiliation{Physics Faculty, Moscow State University, Moscow
119899, Russia}

\author{A.~A.~Cherechukin}
\affiliation{Institute of Radioengineering and Electronics of RAS,
Moscow 103907, Russia}

\author{M.~Matsumoto}
\affiliation{Institute of Multidisciplinary Research for Advanced
Materials, Tohoku University, Sendai 980--8577, Japan}

\author{R.~Note}
\affiliation{Institute for Material Research, Tohoku University,
Sendai 980--8577, Japan}

\begin{abstract}
Polycrystalline $\mathrm{Ni}_{2.18}\mathrm{Mn}_{0.82}\mathrm{Ga}$
Heusler alloys with a coupled magnetostructural transition are
studied by differential scanning calorimetry, magnetic and
resistivity measurements. Coupling of the magnetic and structural
subsystems results in unusual magnetic features of the alloy.
These uncommon magnetic properties of
$\mathrm{Ni}_{2.18}\mathrm{Mn}_{0.82}\mathrm{Ga}$ are attributed
to the first-order structural transition from a tetragonal
ferromagnetic to a cubic paramagnetic phase.
\end{abstract}

\pacs{64.70.Kb, 75.50.Cc}


\maketitle

In $\mathrm{Ni}_{2}\mathrm{MnGa}$, as in many other Heusler alloys
containing manganese, the indirect exchange interaction between
magnetic ions results in ferromagnetism, which is usually
described in terms of the local magnetic moment at the Mn
site.~\cite{plog,kub} This makes it possible to discuss the
occurrence of ferromagnetism in terms of the dependence of the
exchange interaction on the spatial separations of neighboring
manganese atoms. Hydrostatic pressure experiments performed on
ferromagnetic Mn-containing Heusler alloys (see
Refs.~\onlinecite{kano} and \onlinecite{kyuji}, and references
therein) demonstrated that this actually is the case. In contrast
with other Mn-containing Heusler alloys, stoichiometric and
nonstoichiometric Ni-Mn-Ga alloys have been the subjects of
numerous investigations in resent years due to their interesting
physical
properties.~\cite{worg,planes,zhel,cast,ull,tickle,mur,dik}

For stoichiometric $\mathrm{Ni}_{2}\mathrm{MnGa}$, a structural
phase transition of the martensitic type from the parent cubic to
a complex tetragonally based structure takes place at $T_m =
202$~K, whereas ferromagnetic ordering sets at $T_C = 376$~K
(Ref.~\onlinecite{web}). The structural phase transition is driven
by a band Jahn - Teller distortion,~\cite{fujii,brown} and is
accompanied by a reduction in the unit-cell volume. In
Ref.~\onlinecite{vas} it was suggested that in
$\mathrm{Ni}_{2+x}\mathrm{Mn}_{1-x}\mathrm{Ga}$ alloys the
martensitic and ferromagnetic phase transitions couple (i.e., the
order-disorder magnetic transition occurs simultaneously with the
order-order crystallographic phase transition) in the range of
compositions $x = 0.18 - 0.20$.

The phenomenon of coupled magnetic and structural phase
transitions is very rare in condensed-matter physics and just a
few intermetallic compounds with this specific type of transition
have been mentioned in the literature
(Refs.~\onlinecite{levin,roz,zach}, and references therein).
Generally, in this case the strong interrelation of magnetic and
structural subsystems leads to an unusual magnetic behavior of the
materials.~\cite{levin,roz} Such systems, with a close relation
between crystallographic structure and magnetism, are also of a
great technological significance, since they demonstrate such
attractive properties as giant magnetocaloric effect,
magnetostriction, and magnetoresistance.~\cite{VKP,morellon}
Therefore, finding and investigating a system with coupled
magnetostructural transition may be of importance. Although
$\mathrm{Ni}_{2+x}\mathrm{Mn}_{1-x}\mathrm{Ga}$ alloys were
studied by a variety of methods, a systematic study of their
magnetic properties in the vicinity of the magnetostructural
transition is lacking. In this paper we report an experimental
study of the magnetostructural transition in
$\mathrm{Ni}_{2.18}\mathrm{Mn}_{0.82}\mathrm{Ga}$.

A polycrystalline
$\mathrm{Ni}_{2.18}\mathrm{Mn}_{0.82}\mathrm{Ga}$ ingot was
prepared by a conventional arc-melting method in argon atmosphere.
The ingot was homogenized at 1050~K for nine days, and quenched in
ice water. Determined by x-ray diffraction the crystal structure
of the alloy at room temperature has a simple tetragonal
modification with lattice parameters $a = b = 0.774$ nm, and $c =
0.6485$ nm. Samples for differential scanning calorimetry (DSC),
resistivity, and magnetic measurements were spark cut from the
middle part of the ingot. The DSC measurements were done by a
Perkin-Elmer thermal analysis equipment with a heating/cooling
rate of 5~K/min. Temperature and magnetic field dependencies of
magnetization were performed by a superconducting quantum
interference device magnetometer. Electrical resistivity was
measured by a standard four-probe technique in a zero magnetic
field with a heating/cooling rate of 1~K/min.

As evident from the DSC measurements (Fig.~1), both direct and
reverse martensitic transitions are accompanied by a well-defined
peak on the heat flow due to the latent heat of the transition.
Critical temperatures of the martensitic transition, austenite
finish ($A_f$), and martensite start ($M_s$), determined from DSC
curves are equal to $A_f = 338$~K and $M_s = 330$~K. A temperature
hysteresis of the martensitic transformation estimated as $\delta
T = A_f - M_s$ was found to be 8~K. The calculated value of the
latent heat of transition, $Q = 9.6$~J/g, is in agreement with
values obtained for nonstoichiometric alloys by Chernenko
\textit{et al}.~\cite{cher}

\begin{figure}
\includegraphics[width=\columnwidth]{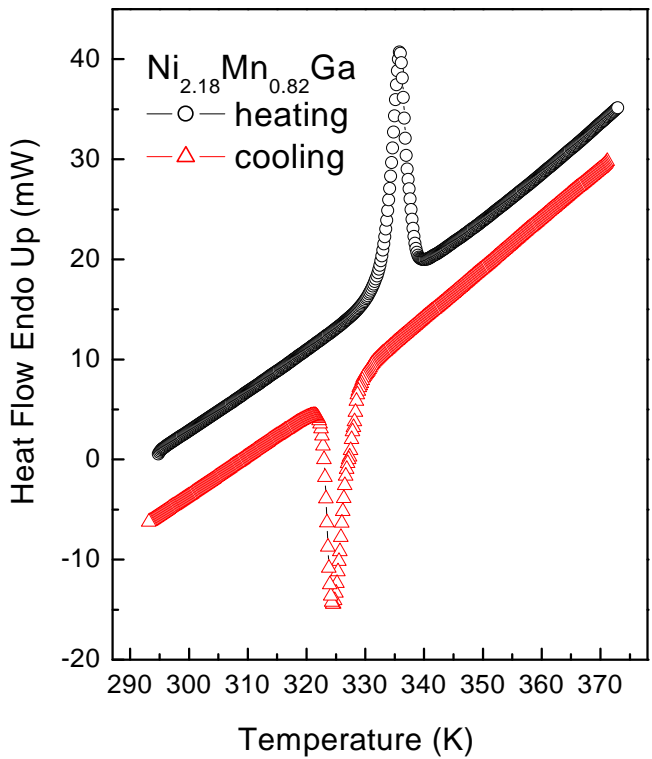}
\caption{\label{fig-1} Heat flow measured during heating and
cooling for $\mathrm{Ni}_{2.18}\mathrm{Mn}_{0.82}\mathrm{Ga}$.}
\end{figure}

\begin{figure}
\includegraphics[width=\columnwidth]{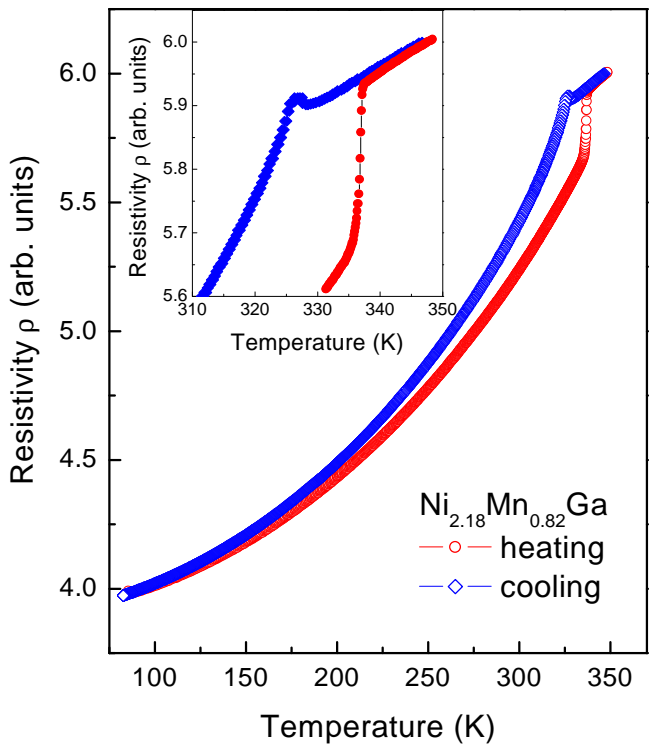}
\caption{\label{fig-2} Temperature dependencies of the electrical
resistivity for $\mathrm{Ni}_{2.18}\mathrm{Mn}_{0.82}\mathrm{Ga}$
measured during heating and cooling. The inset shows the behavior
of the resistivity in the vicinity of the magnetostructural
transition in more detail.}
\end{figure}

The temperature dependencies of electrical resistivity of
$\mathrm{Ni}_{2.18}\mathrm{Mn}_{0.82}\mathrm{Ga}$, during heating
and cooling, are shown in Fig.~2. Generally, the resistivity has a
normal metallic character, and upon heating continuously increases
due to the increase in both electron-phonon and electron-magnon
scatterings. A distinguishing feature of the resistivity is a very
large temperature hysteresis between heating and cooling
processes, extending up to several hundreds degrees. The behavior
of the resistivity in the neighborhood of the magnetostructural
transition (inset in Fig.~2) gives rise to the conclusion that the
transition has a temperature hysteresis. Indeed, in ferromagnetic
materials $\rho(T)$ can be presented as $\rho(T) = \rho_{0} +
\rho_{ph} + \rho_{mag}$, where $\rho_0$ represents the residual
resistivity, and $\rho_{ph}$ and $\rho_{mag}$ are contributions
due to the lattice vibration and spin-disorder scattering,
respectively. In a paramagnetic state $\rho_{mag}$ becomes
temperature independent, and the resistivity increases due to the
contribution from electron-phonon scattering, which can be
approximated as $\rho_{ph} = AT$, with $A$ a constant. As seen
from Fig.~2, for the alloy studied the linear part of the
resistivity starts at 337~K upon heating, but extends down to
329~K upon subsequent cooling. These critical temperatures, 337
and 329~K, are in good consistent with $A_f$ and $M_s$
temperatures determined from the DSC measurements (Fig.~1).

\begin{figure}
\includegraphics[width=\columnwidth]{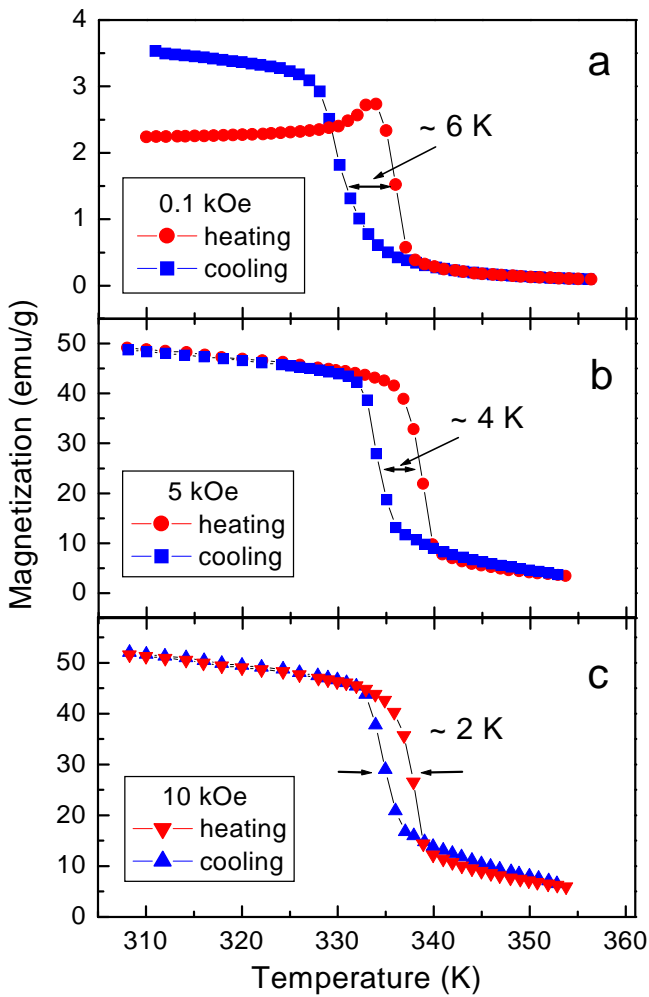}
\caption{\label{fig-3} The temperature dependencies of
magnetization of $\mathrm{Ni}_{2.18}\mathrm{Mn}_{0.82}\mathrm{Ga}$
during continuous heating and cooling in various magnetic fields.}
\end{figure}

\begin{figure}
\includegraphics[width=\columnwidth]{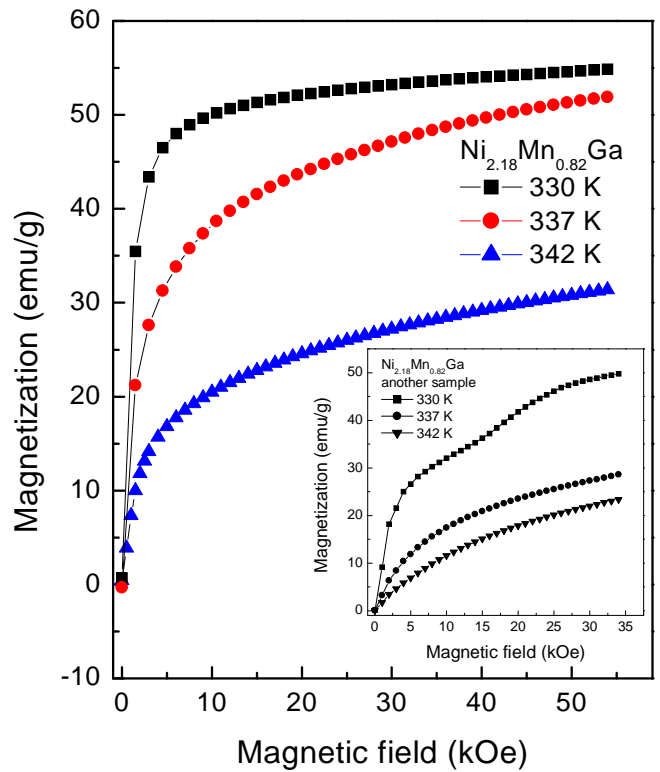}
\caption{\label{fig-4} Magnetic field dependence of the
magnetization $M$ of
$\mathrm{Ni}_{2.18}\mathrm{Mn}_{0.82}\mathrm{Ga}$ measured in the
vicinity of the phase transition. The inset shows the results of
$M(H)$ measured for another
$\mathrm{Ni}_{2.18}\mathrm{Mn}_{0.82}\mathrm{Ga}$ sample.}
\end{figure}

The temperature dependencies of the magnetization $M$ of
$\mathrm{Ni}_{2.18}\mathrm{Mn}_{0.82}\mathrm{Ga}$, for continuous
heating and cooling, confirm the hysteretic feature of the
magnetostructural transition (Fig.~3). The temperature hysteresis
of magnetization in a 0.1~kOe magnetic field [Fig.~3(a)] is
estimated as $\delta T \simeq 6$~K, in good accordance with the
results of DSC and resistivity measurements. A notable feature of
the magnetization measured in this field is that during heating
$M$ increases up to a temperature at which the transformation into
a paramagnetic state occurs. We suggest that this behavior is
caused by the formation of ferromagnetic austenite, which
possesses a higher magnetization in low magnetic
fields.~\cite{web} Measurements of the magnetization in higher
magnetic fields [Figs.~3(b) and~3(c)] point to a shift of the
characteristic temperatures of the magnetostructural transition,
which is accompanied by a narrowing of the temperature hysteresis
of magnetization. The critical temperatures of the direct and
reverse martensitic transitions $A_f$ and $M_s$, determined as a
minimum on the $(\partial M/\partial T)_H$ derivative, were found
to increase from 336 to 338~K for $A_f$ temperature and from 330
to 335~K for $M_s$ temperature as the magnetic field increases
from 0.1 to 10~kOe.

Since representatives of intermetallic compounds with coupled
magnetostructural transitions usually show pronounced anomalies on
isothermal magnetization curves,~\cite{levin} a similar behavior
could be expected for
$\mathrm{Ni}_{2.18}\mathrm{Mn}_{0.82}\mathrm{Ga}$. With this aim
we undertook measurements of isothermal magnetization in the
vicinity of the magnetostructural transition. Contrary to the
expectation, the measurements revealed field dependencies of the
magnetization, typical of a ferromagnet (Fig.~4). At $T = 330$~K,
the magnetization saturates in a magnetic field of 5~kOe. The
value of the magnetization saturation, $M_s \simeq 54$~emu/g, is
in good correspondence with the results of $M(T)$ measurements
[Fig.~3(b)]. However, the $M(H)$ dependencies measured for another
Ni$_{2.18}$Mn$_{0.82}$Ga sample at the same temperatures revealed
that the isothermal magnetization process is essentially sample
dependent. As is seen from the inset in Fig.~4, a well-defined
anomaly appears at the $M(H)$ curve taken at 330~K when the
magnetic field reaches a critical value of 15~kOe. The $M(H)$
dependencies, measured at higher temperatures, show a monotonic
increase of $M$; thus no anomaly is seen up to the maximum
magnetic field used in the measurements. Since
Ni$_{2.18}$Mn$_{0.82}$Ga is on the edge of the compositional
interval of alloys with coupled magnetostructural transitions,
this drastic difference in the magnetization process between two
sample could be due to a slight compositional inhomogeneity which
is intrinsic for intermetallic compounds.

The spatial dependence of the Mn - Mn exchange interaction in
Mn-containing Heusler alloys was demonstrated by hydrostatic
pressure experiments.~\cite{kano,kyuji} The authors found that the
Curie temperature $T_C$ of $\mathrm{Ni}_{2}\mathrm{MnZ}$ (Z = Al,
Ga, In, Sn, and Sb) increases with increasing pressure. Based on
these and previously published results, Kanomata \textit{et
al.}~\cite{kano} defined an interaction curve where $T_C$ was
plotted as a function of the interatomic distance of Mn-Mn atoms.
This curve also supports the idea that ferromagnetic ordering in
Heusler alloys containing manganese depends strongly and
definitely on the Mn-Mn distance.

This strong dependence of $T_C$ on Mn-Mn distance is presumably
responsible for the surprising temperature hysteresis in the
ferromagnetic ordering of
$\mathrm{Ni}_{2.18}\mathrm{Mn}_{0.82}\mathrm{Ga}$ observed by the
resistivity (Fig.~2) and confirmed by the magnetization
measurements (Fig.~3). Since previous magnetic studies of Ni-Mn-Ga
did not report on a hysteretic feature of
magnetization,~\cite{vas-1990,hu} it can be assumed that the
coupling of magnetic and structural transitions plays the sole
role in the hysteretic behavior of the magnetization in
Ni$_{2.18}$Mn$_{0.82}$Ga. Taking into account the crucial role of
the nearest Mn-Mn distance in the formation of long-range
ferromagnetic ordering, a first-order martensitic transition from
a low-temperature low-volume tetragonal phase to a
high-temperature high-volume cubic phase provokes a temperature
hysteresis of the magnetization due to the fact that the
temperature at which the high-volume cubic phase nucleates upon
warming up is higher than the temperature at which it transforms
to the low-volume tetragonal phase upon subsequent cooling down.
Comparison of the results of the DSC and resistivity measurements
(Figs.~1 and~2) reflects this strong sensitivity of the long-range
ferromagnetic order to the unit-cell volume. As is evident from
Fig.~2, during warming up the sample transforms to the
paramagnetic state at 337~K, which matches with the austenite
finish temperature $A_f$ = 338~K determined from the DSC
measurements. Upon subsequent cooling down the linear temperature
dependence of the resistivity, suggesting that the sample is in
paramagnetic state, extends down to 329~K, which agrees well with
the onset of formation of the low-volume martensitic phase: $M_s$
= 330~K. Therefore, from these results it can be deduced that the
ferromagnetic ordering sets in (or disappears) not by means of a
classical mechanism but due to the structural transformation.

The shift of transition temperature in a magnetic field can be
estimated using the Clapeyron-Clausius equation. For a
transformation from ferromagnetic martensite to paramagnetic
austenite, the temperature shift has the form $\Delta T =
M_MHT_m/Q$ (Ref.~\onlinecite{vas}), where $M_M$, $H$, $T_m$, and
$Q$ are magnetization of martensite, magnetic field, martensitic
transition temperature, and the latent heat of transition,
respectively. From our experimental results $M_s \simeq 50$~emu/g,
$Q$ = 9.6~J/g, and $T_m = (A_f + M_s)/2 = 333$~K, and the
calculation yields $d\Delta T/dH \simeq 0.18$~K/kOe. This
estimation is less than half the experimental value determined
from the magnetization data, $\sim 0.35$~K/kOe. This experimental
value as well as the observed narrowing of the hysteresis interval
is in contradiction with the experimental results obtained by an
optical method~\cite{bozhko} for Ni$_{2.19}$Mn$_{0.81}$Ga in
magnetic fields up to 10~kOe, which suggest that the temperatures
of the direct and reverse martensitic transformations as functions
of the applied magnetic field, have a linear field dependence with
a slope of $\simeq 0.15$~K/kOe. These differences between
Ni$_{2.18}$Mn$_{0.82}$Ga and Ni$_{2.19}$Mn$_{0.81}$Ga can comprise
evidence that in the interaction of magnetic and structural
subsystems in Ni$_{2+x}$Mn$_{1-x}$Ga ($x = 0.18 - 0.20$), chemical
composition is the crucial factor.

Theoretical calculations~\cite{dik} for Ni$_{2+x}$Mn$_{1-x}$Ga
with $x \leq 0.16$ indicate that generally the temperature of the
direct martensitic transformation is more sensitive to the
application of a magnetic field than that of the reverse
martensitic transformation. When compared to the theoretical
results,~\cite{dik} the field dependencies of the transition
temperatures determined from our experiment have anomalously high
rates. This can be accounted for by the fact that, contrary to the
$x \leq 0.16$ alloys, where a martensitic transformation occurs in
the ferromagnetic matrix, in the alloy studied the structural
transformation also results in a switch of the magnetic state of
the material.

It was verified experimentally for
$\mathrm{Ni}_{2+x}\mathrm{Mn}_{1-x}\mathrm{Ga}$
alloys~\cite{dik,bozhko,inoue} that high magnetic fields favor a
martensitic phase. This suggests that if, at some temperature
fixed within an interval from $A_s$ to $A_f$, there is a fraction
of paramagnetic austenite, it will be transformed to a highly
magnetized state at some critical value of the applied magnetic
field. Therefore, it is likely that the anomaly observed at the
magnetization isotherm taken at 330~K (the inset in Fig.~4) is due
to a magnetic-field-induced transformation of a fraction of
austenite which is paramagnetic at the given temperature to the
ferromagnetic martensite. Since such a transformation results in
marked anomalies on isothermal magnetization of
$\mathrm{Gd}_{5}(\mathrm{Si}_{x}\mathrm{Ge}_{4-x})$
alloys,~\cite{levin} more pronounced anomalies can be observed at
$M(H)$ for $\mathrm{Ni}_{2+x}\mathrm{Mn}_{1-x}\mathrm{Ga}$ $(x =
0.18 - 0.20)$ at higher temperatures and in stronger magnetic
fields.

In summary, in this paper we have reported on unusual magnetic
behavior (first-order character of the ferromagnetic -
paramagnetic transition) in
$\mathrm{Ni}_{2.18}\mathrm{Mn}_{0.82}\mathrm{Ga}$ Heusler alloy.
The obtained experimental results imply that the uncommon magnetic
properties of the alloy originate from a first-order structural
transition from a tetragonal ferromagnetic structure to a cubic
paramagnetic structure. The observed narrowing of the temperature
hysteresis of magnetization at high magnetic fields is attributed
to differences in the magnetic field dependencies of the
temperatures of the direct and reverse martensitic
transformations. Since magnetostructural transitions frequently
lead to physical properties of technological significance, our
finding may also make the Ni$_{2+x}$Mn$_{1-x}$Ga ($x = 0.18 -
0.20$) alloys attractive from the point of view of their potential
application for magnetic refrigeration or magnetostrictive
transducers.

\medskip

We are grateful to Professor A.~N.~Vasil'ev for helpful
discussions. Christina Wedel is acknowledged for help with x-ray
diffraction measurements. This work was partially supported by the
Grant-in-Aid for Scientific Research (C) No. 11695038 from the
Japan Society of the Promotion of Science and by a Grant-in-Aid of
the Russian Foundation for Basic Research No. 99-02-18247.

\end{document}